\documentclass[aps,pra,twocolumn,showpacs]{revtex4}
\usepackage{graphicx}
\usepackage{amsmath}
\usepackage{dcolumn}
\usepackage{bm}

\newcommand{\beq}{\begin{equation}}
\newcommand{\eeq}{\end{equation}}
\newcommand{\beqa}{\begin{eqnarray}}
\newcommand{\eeqa}{\end{eqnarray}}

\def\ajp#1{{ Am.\ J.\ Phys.} {\bf #1}}

\def\ol#1{{ Opt.\ Lett.} {\bf#1}}
\def\jmo#1{{ J.\ Mod.\ Opt.} {\bf#1}}
\def\jpb#1{{ J.\ Phys.\ B} {\bf#1}}

\def\nat#1{{ Nature} {\bf#1}}

\def\phrep#1{{ Phys.\ Reports\/} {\bf#1}}
\def\phtod#1{{ Phys.\ Today} {\bf#1}}

\def\pra#1{{ Phys.\ Rev. A\/} {\bf#1}}
\def\prb#1{{ Phys.\ Rev. B\/} {\bf#1}}
\def\prd#1{{ Phys.\ Rev. D\/} {\bf#1}}

\def\prl#1{{ Phys.\ Rev.\ Lett.} {\bf#1}}

\def\sci#1{{ Science} {\bf#1}}
\def\rmp#1{{ Rev. \ Mod. \ Phys.} {\bf#1}}

\begin{document}

\title{Nonlinear Constants of Quantum Information in Reversible and Irreversible Amplitude Flows}
\author{Xiao-Feng Qian and J.H. Eberly}
\affiliation{Rochester Theory Center and Department of Physics \& Astronomy\\
University of Rochester, Rochester, New York 14627}
\date{\today }

\begin{abstract}
We report an approach to quantum open system dynamics that leads to
novel nonlinear constant relations governing information flow among
the participants. Our treatment is for mixed state systems entangled
in a pure state fashion with an unspecified party that was involved
in preparing the system for an experimental test, but no longer
interacts after $t=0$. Evolution due to subsequent interaction with
another party is treated as an amplitude flow channel and uses
Schmidt-type bipartite decomposition of the evolving state. We
illustrate this with three examples, including both reversible and
irreversible information flows, and give formulas for the new
nonlinear constraints in each case.
\end{abstract}


\maketitle

%+++++++++++++++++++++++++++++++++++++++++++++
\section{Introduction}
Entanglement, as a term of joint quantum coherence, is one of the
most intriguing elements of quantum mechanics and it is crucial in
quantum information tasks \cite{NC-Preskill}. However the existence
of an interacting reservoir or environment that leads to decoherence
and/or disentanglement \cite{Zurek91, Chuang-etal95, Yu-Eberly04}
places an obstacle to the maintenance of joint quantum coherence
during any dynamical process \cite{maintain coherence}. Thus the
study and control of entanglement dynamics has received wide
attention in recent years (see reviews \cite{Mintert-etal05,
Amico-etal08, Yu-EberlySci09}). There have been studies of
entanglement dynamics from many points of view. Examples involve
open system treatments \cite{Nha-Carmichael04, ESD-early, ESD-later,
ESB, Kim-etal} or closed quantum scenarios such as cavity QED
systems \cite{Phoenix-Knight, Bose-etal01, Yonac-etal06,
Yonac-etal07, Son-etal02, Sainz-Bjork07, Yonac-Eberly08,
Yonac-Eberly10}, spin systems \cite{Subrahmanyam04, Qian-etal,
Cubitt-etal, Chan-etal08, Wang01, Pratt-Eberly01, Amico-etal04,
Yang-etal06}, etc.

Many interesting and sometimes surprising findings such as
entanglement sudden death \cite{Yu-Eberly04, ESD-early, ESD-later},
sudden birth \cite{ESB}, revivals \cite{Kim-etal, Phoenix-Knight,
Yonac-Eberly08}, dynamical relations with quantum state transfer
\cite{Subrahmanyam04, Qian-etal, Yang-etal06}, and other exotic
types of entanglement evolution have been reported. Such interesting
phenomena accompany the idea of tracking entanglement as a carrier
of quantum information \cite{ESD-later, ESB, Kim-etal,
Phoenix-Knight, Bose-etal01, Yonac-etal06, Yonac-etal07, Son-etal02,
Sainz-Bjork07, Yonac-Eberly08, Yonac-Eberly10, Subrahmanyam04,
Qian-etal, Yang-etal06, Cubitt-etal, Chan-etal08}, a generalization
of entanglement swapping \cite{swapping}.

One consequence has been the discovery of examples of non-trivial
``information conservation" among three or more parties
\cite{Yonac-etal07, Sainz-Bjork07, Chan-etal08}, arising in cases of
sufficiently symmetric interaction Hamiltonians, or special initial
states, or reservoirs that are sufficiently small that their state
evolution can be followed in detail, such as in perfect-mirror
closed-system cavity QED \cite{Yonac-etal06, Yonac-etal07,
Sainz-Bjork07, Yonac-Eberly08, Yonac-Eberly10} and in spin systems
\cite{Qian-etal, Yang-etal06, Cubitt-etal, Chan-etal08}. However
true reservoirs are complex and difficult to follow, especially if
mixed state considerations are important. No general closed-form
rules of entanglement transfer are known in such cases.

In this paper we revisit quantum information flow from a different
perspective and derive a new class of entanglement constants of
motion. Our approach employs amplitude channel dynamics and avoids
information loss by tracing, while remaining open to non-Markovian
as well as Markovian reservoir behavior. We note that the system of
experimental interest, which may be one or more qubits, is almost
always prepared in a pure state if possible, and frequently the
method of preparation produces a pure entangled state. This means
that the system itself is in a mixed rather than a pure state. We
assume that the entanglement during state preparation, causing the
mixedness, arose via interactions that have ceased prior to the
beginning of a period of interest at $t = 0$. This period of
interest could simply be intended for quantum memory preservation or
for specific state manipulations. The static disengaged nature of
the prior entanglement partner, and also its lack of specificity in
our treatment, reduce it to a vague background object in any further
qubit evolution, and for this reason we label it the ``Moon".

\begin{figure}[b!]
\includegraphics[width=4cm]{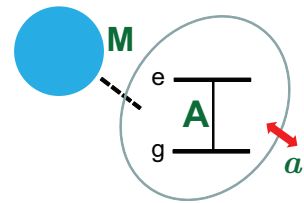}
\caption{A general sketch of our scenario. The bubble circles the
system of interest $A$ and leaves everything else out. The dashed
line indicates its entanglement, but not interaction, with the
unspecified background ``Moon" $M$. The arrow represents interaction
between $A$ and an arbitrarily-dimensioned unit $a$, which can be
the quantum vacuum reservoir, a single mode cavity, an XY spin
chain, etc.} \label{Moon}
\end{figure}

A general sketch of our scenario is given in Fig. \ref{Moon}. Unit
$A$ is taken as a two-level system (qubit) and unit $a$ as a
separate quantum system of arbitrary dimension interacting with it,
nominally a reservoir. The Moon $M$, i.e., the non-interacting,
unspecified, and completely static background, is entangled via an
earlier preparation stage with $A$.

There obviously remains a wide choice for systems acting as
environments that promote evolution of the system of interest after
$t = 0$.  We will illustrate a range of possibilities with concrete
results in various specific interaction contexts: spontaneous
emission \cite{W-W}, Jaynes-Cummings (JC) cavity dynamics \cite{JC},
and XY spin chain interactions \cite{Lieb61}. These present very
different physical situations and interaction mechanisms, and lead
to distinct entanglement dynamics, but they all react similarly to
the initial Moon entanglement. Our linked information constants
arise from amplitude channel dynamics but do not rely on symmetries
of the Hamiltonian or of any special initial state, in contrast to
the cases in some previous work \cite{Yonac-etal07, Chan-etal08}.

%+++++++++++++++++++++++++++++++++++++++++++++
\section{Schmidt analysis of Entanglement}

In this section we address our approach to entangled state analysis.
The Hamiltonian of our scheme reads
\begin{equation}
H=H_{A}+H_{a}+H_{Aa}+H_{M},
\end{equation}
where $H_{A}$, $H_{a}$ and $H_{M}$ are the Hamiltonians of the qubit
system $A$, ``reservoir" unit $a$ and the previously-interacting
Moon $M$ respectively; and $H_{Aa}$ denotes the only existing
interaction, that between $A$ and $a$.

We start from the $A$-$M$ entangled preparation state, i.e., the
joint superposition state
\begin{equation}
|\psi _{AM}(0)\rangle =\cos \theta |e\rangle |m_{1}\rangle +\sin
\theta |g\rangle |m_{2}\rangle, \label{joint}
\end{equation}
where $|e\rangle $ and $|g\rangle $ are the excited and ground
states of our qubit system $A$, and
\begin{equation}
|m_{1}\rangle =\sum_{l}u_{l}|M_{l}\rangle, \quad \text{and} \quad
|m_{2}\rangle =\sum_{l}v_{l}|M_{l}\rangle,
\end{equation}
are two normalized Moon states with $\{|M_{l}\rangle \}$ defined as
a complete basis set for $M$. It need not be the case generally, but
we assume in our example that the Moon states $|m_{1}\rangle $ and
$|m_{2}\rangle $ are orthogonal, and because $M$ is not interacting
with either $A$ or $a$ they are effectively static:
\begin{equation}
\langle m_{2}(t)|m_{1}(t)\rangle =\langle m_{2}|m_{1}\rangle =0.
\end{equation}

We adopt a conventional approach to the interacting partner system
labelled $a$, assuming that it is separable from the qubit $A$ at $t
= 0$, just as in the conventional treatment of a reservoir in
quantum open system dynamics \cite{Zurek91, Chuang-etal95,
Yu-Eberly04}. Therefore the entire initial state can be written as
\begin{equation}
|\psi _{AaM}(0)\rangle =\Big(\cos \theta |e\rangle |m_{1}\rangle
+\sin \theta |g\rangle |m_{2}\rangle \Big)\otimes |\phi_{1}\rangle,
\label{initial}
\end{equation}
where $|\phi _{1}\rangle =\sum_{k}i_{k}|k\rangle$ is a normalized
state of unit $a$, and $\{|k\rangle\}$ is a complete basis for $a$.
Usually $|\phi _{1}\rangle$ is the ground state of part $a$. We note
that since $M$ is not interacting, the evolution of the states
$|e\rangle |\phi _{1}\rangle $ and $|g\rangle |\phi _{1}\rangle $
are driven only by the Hamiltonian $H_{Aa}$, i.e., the dynamics of
the $A$-$a$ part can be separated from $M$. Therefore we will only
need to focus on the $A$-$a$ dynamics when we study the time
evolution.

Before we proceed to the time dependent state in various specific
models in the following sections, we will first discuss the initial
Moon entanglement. As we know, the pure-state relation between the
two sides of any bi-partition is an $R\times S$ dimensional matrix
(where $R$ and $S$ can be any numbers or infinite) that may connote
entanglement, but in any event permits a Schmidt-type decomposition
of the joint state \cite{Grobe-etal, Ekert, EberlyLa06}. We use the
Schmidt parameter $K$ introduced by Grobe, et al., \cite{Grobe-etal}
as our quantitative measure of entanglement, where $K$ is not simply
the dimension of the space \cite{NC-Preskill} but rather relates to
the number of Schmidt modes that make a significant contribution to
the state. Therefore we name this parameter $K$ the ``Schmidt
weight" from now on. The range of this Schmidt weight, $N \ge K \ge
1$ ($N$ is the effective dimension of the space) corresponds to the
concurrence range $1 \ge \mathcal{C} \ge 0$, when concurrence
\cite{Wootters98} is also applicable. The upper and lower ends of
both ranges denote maximal and zero entanglement, respectively.

The Schmidt weight between two parties $\alpha$ and $\beta $ of a
general pure state $|\psi_{\alpha\beta}\rangle$ is defined as
\begin{equation}
K=\Big[{\sum_{k}\lambda_{k}^{2}}\Big]^{-1},  \label{Schmidt}
\end{equation}
where these $\lambda_{k}$s are the non-zero eigenvalues of the
reduced density matrix for either system, $\rho_{\alpha }$ or
$\rho_{\beta}$ \cite{EberlyLa06}:
\begin{equation}
\rho_{\beta}=\mathrm{Tr}_{\alpha}[\rho]=\mathrm{Tr}_{\alpha}
\Big[|\psi_{\alpha\beta}\rangle\langle\psi_{\alpha\beta}|\Big]
=CC^{\dag}.
\end{equation}
Here $C$ is the coefficient matrix connecting the two separate
arbitrary complete bases $|n\rangle $ and $|\mu\rangle$ of systems
$\alpha$ and $\beta$ respectively, with
\begin{equation}
|\psi_{\alpha\beta}\rangle=\sum_{n,\mu}C(n,\mu)|n_{\alpha}\rangle\otimes
|\mu_{\beta}\rangle.  \label{psi alphabeta}
\end{equation}
The square roots of $\lambda_{k}$s are also the coefficients of the
usual Schmidt decomposition \cite{NC-Preskill}
\begin{equation}
|\psi_{\alpha\beta}\rangle=\sum_{k}\sqrt{\lambda_{k}}|f_{k}^{\alpha}\rangle
\otimes |g_{k}^{\beta}\rangle,
\end{equation}
where $|f_{k}^{\alpha}\rangle$ and $|g_{k}^{\beta}\rangle$ are the
orthonormal Schmidt states satisfying $\langle
f_{k}^{\alpha}|f_{k'}^{\alpha}\rangle=\langle
g_{k}^{\beta}|g_{k'}^{\beta}\rangle=\delta_{kk'}$.

Since a general pure state is usually in some arbitrary basis other
than the Schmidt basis, it is natural for us to follow the
coefficient matrix procedure to calculate the Schmidt weight
(\ref{Schmidt}). Accordingly we note from the initial state
(\ref{initial}) that the coefficient matrix for the Moon $M$ in the
basis of $|m_{1}\rangle$, $|m_{2}\rangle $, $|m_{3}\rangle$, ...,
and the interacting partner $a$ in the basis $\{|k\rangle\}$, is an
$\infty \times \infty $ matrix which is given as
\begin{equation}
C_{M}=\left(
\begin{array}{cccccccc}
i_{1}\cos \theta  & .. & i_{k}\cos \theta  & .. & 0 & ... & 0 & ... \\
0 & ... & 0 & ... & i_{1}\sin \theta  & .. & i_{k}\sin \theta  & .. \\
\vdots  & \vdots  & \vdots  & \vdots  & \vdots  & \vdots  & \vdots &
\vdots
\end{array}
\right),
\end{equation}
where the two-dot sign ``$..$" represents the elements
$i_{k}\cos\theta$ for all the other $k$s, while the three-dot sign
``$...$" represents empty rows or columns of zeros for the infinite
number of remaining matrix elements. The reduced density matrix
however is simply
\begin{equation}
\rho _{M}=C_{M}C_{M}^{\dag}=\left(
\begin{array}{ccc}
\cos ^{2}\theta  & 0 & \cdots  \\
0 & \sin ^{2}\theta  & \cdots  \\
\vdots  & \vdots  & \ddots
\end{array}
\right) .
\end{equation}

We note that the qubit has reduced the effective interaction space
of the Moon to a two dimensional subspace, which means that in this
context a two-state subspace of the Moon is in fact quite general.
The non-zero eigenvalues of the above matrix are obvious, and the
resulting Schmidt weight $K_{M}$ denoting the entanglement between
the Moon the remainder is given as
\begin{equation}
K_{M} = \frac{1}{\cos^{4}\theta + \sin^{4}\theta}.
\label{MoonEntanglement}
\end{equation}

Since the Moon $M$ is not interacting, its internal dynamics only
amount to a local unitary transformation \cite{Nielsen99}, which
will not affect the entanglement between $M$ and the rest, so $K_M$
is independent of time. We note that there is no Moon entanglement
($K_{M}=1$) when $\theta = \pi/2$ or 0. In these cases the initial
state is a trivial product state. Otherwise $K_{M}>1$. It is
particularly interesting when the Moon restricts the $A$-$a$
dynamics and acts as a monitor of the entire entanglement flow. The
following sections take a few specific examples of the $A$-$a$
interaction to show the role of Moon entanglement in their
particular entanglement information dynamics.

%+++++++++++++++++++++++++++++++++++++++++++++
\section{Spontaneous Emission}

In this case the qubit system $A$ is a two level atom and unit $a$
is the quantum vacuum reservoir consisting of the continuum of
photon modes. The atom will of course decay to its ground state
asymptotically and irreversibly, while one photon is emitted
\cite{W-W, Def-gk}. We write the Hamiltonian in the usual way as a
sum of atom and reservoir contributions:
\begin{equation}
H_{A}=\frac{1}{2}\hbar \omega _{A}\sigma _{A}^{z}\quad
\mathrm{and}\quad H_{a}=\sum_{k}\hbar \omega _{k}a_{k}^{\dag }a_{k}.
\end{equation}
Here $\sigma_{A}^{z}$ is the usual Pauli matrix, and the usual boson
operators represent the reservoir with a continuum of modes, where
$a_{k}^{\dag }$ and $a_{k}$ denote the standard creation and
annihilation operators respectively, and $\omega _{A}$ and $\omega
_{k}$ are the atom and reservoir frequencies. Here
$k=1,2,3,...,\infty$, labels the infinitely many modes. The
interaction Hamiltonian is also standard:
\begin{equation}
H_{I}=\sum_{k}\hbar (g_{k}^{\ast}
\sigma_{A}^{-}a_{k}^{\dag}+g_{k}\sigma_{A}^{+}a_{k}),
\end{equation}
where $\sigma_{A}^{+}$, $\sigma _{A}^{-}$ are the usual raising and
lowering Pauli operators for the two level system $A$, and the
$g_{k}$s are coupling constants between the reservoir and the atom,
for which fundamental expressions are well known \cite{Def-gk}
\begin{equation}
|g_{k}|^{2}=\frac{\omega _{k}}{2\hbar \epsilon_{0}V}{\bf
d}_{10}^{2}\cos ^{2}\theta.
\end{equation}
Here $\theta $ is the angle between the atomic dipole moment
$\mathbf{d}_{10} $ and the electric field polarization vector
$\hat{\epsilon}_{k}$, and $V$ is the quantization volume. According
to our generic description in Eq. (\ref{initial}), the initial state
can be rewritten in the spontaneous emission case as
\begin{equation}
|\psi_{AaM}(0)\rangle =\Big(\cos \theta |e\rangle |m_{1}\rangle
+\sin \theta |g\rangle |m_{2}\rangle \Big)\otimes |0\rangle,
\end{equation}
where $|e\rangle$ and $|g\rangle$ are the excited and ground state
of the two level atom, and we have defined $|\phi _{1}\rangle
=|0\rangle$, indicating that all the reservoir modes are in their
vacuum states. With the help of the Weisskopf-Wigner treatment
\cite{W-W, Def-gk} we will find
\begin{eqnarray}
|\psi _{AaM}(t)\rangle  &=&\cos \theta |m_{1}(t)\rangle
\Big(c_{e}(t)|e\rangle |0\rangle +\sum_{k}c_{k}(t)|g\rangle
|1_{k}\rangle \Big) \notag \\
&+& \sin \theta |m_{2}(t)\rangle |g\rangle |0\rangle, \label{Spon
state}
\end{eqnarray}
where the coefficient $c_{e}(t)=e^{-\Gamma _{A}t/2}$\ with
$\Gamma_{A}={\bf d}_{10}^{2}\omega _{A}^{3}/3\pi \epsilon _{0}\hbar
c^{3}$ as the natural line width, $|1_{k}\rangle $ denotes that
there is one photon in the reservoir mode $k$ while all the rest of
the modes are empty and the coefficients
\begin{equation}
c_{k}(t)=g_{k}\frac{1-e^{i(\omega_{A}-\omega
_{k})t-\Gamma_{A}t/2}}{\omega_{k}-\omega_{A}+i\Gamma_{A}t/2}
\end{equation}
are time dependent. The probability to find the atom in the excited
state is $P_{e}(t)=|c_{e}(t)|^{2}=e^{-\Gamma _{A}t}$, which decays
to zero asymptotically and irreversibly.

With the dynamical state (\ref{Spon state}) we can begin to
calculate the Schmidt weight $K_{A}(t)$, or $K_{a}(t)$, representing
the entanglement between qubit $A$, or vacuum reservoir $a$, and
their corresponding remainders. As defined in the last section, the
coefficient matrix between $A$ and the remainder for the time
dependent state (\ref{Spon state}) is given as
\begin{equation}
C_{A}=\left(
\begin{array}{cccccc}
c_{e}(t)\cos \theta  & 0 & 0 & ... & 0 & ... \\
0 & \sin\theta & c_{1}(t)\cos\theta & .. & c_{k}(t)\cos\theta &..
\end{array}
\right),
\end{equation}
where the two-dot sign ``$..$" represents $c_{k}(t)\cos \theta$ for
all the other $k$s and the three-dot sign ``$...$" again represents
empty rows or columns of zeros. Then the reduced density matrix is
simply a $2\times 2 $ form
\begin{equation}
\rho _{A}=\left(
\begin{array}{cc}
|c_{e}(t)|^{2}\cos ^{2}\theta  & 0 \\
0 & \sum_{k}|c_{k}(t)|^{2}\cos ^{2}\theta +\sin ^{2}\theta
\end{array}
\right).
\end{equation}
Now according to the definition in Eq. (\ref{Schmidt}), we
immediately have the qubit entanglement $K_{A}(t)$ given by the
expression
\begin{equation}
K_{A}(t)=\frac{2}{[2\cos ^{2}\theta e^{-\Gamma _{A}t}-1]^{2}+1}.
\label{SponK_A}
\end{equation}

\begin{figure}[t!]
\includegraphics[width= 4.2 cm]{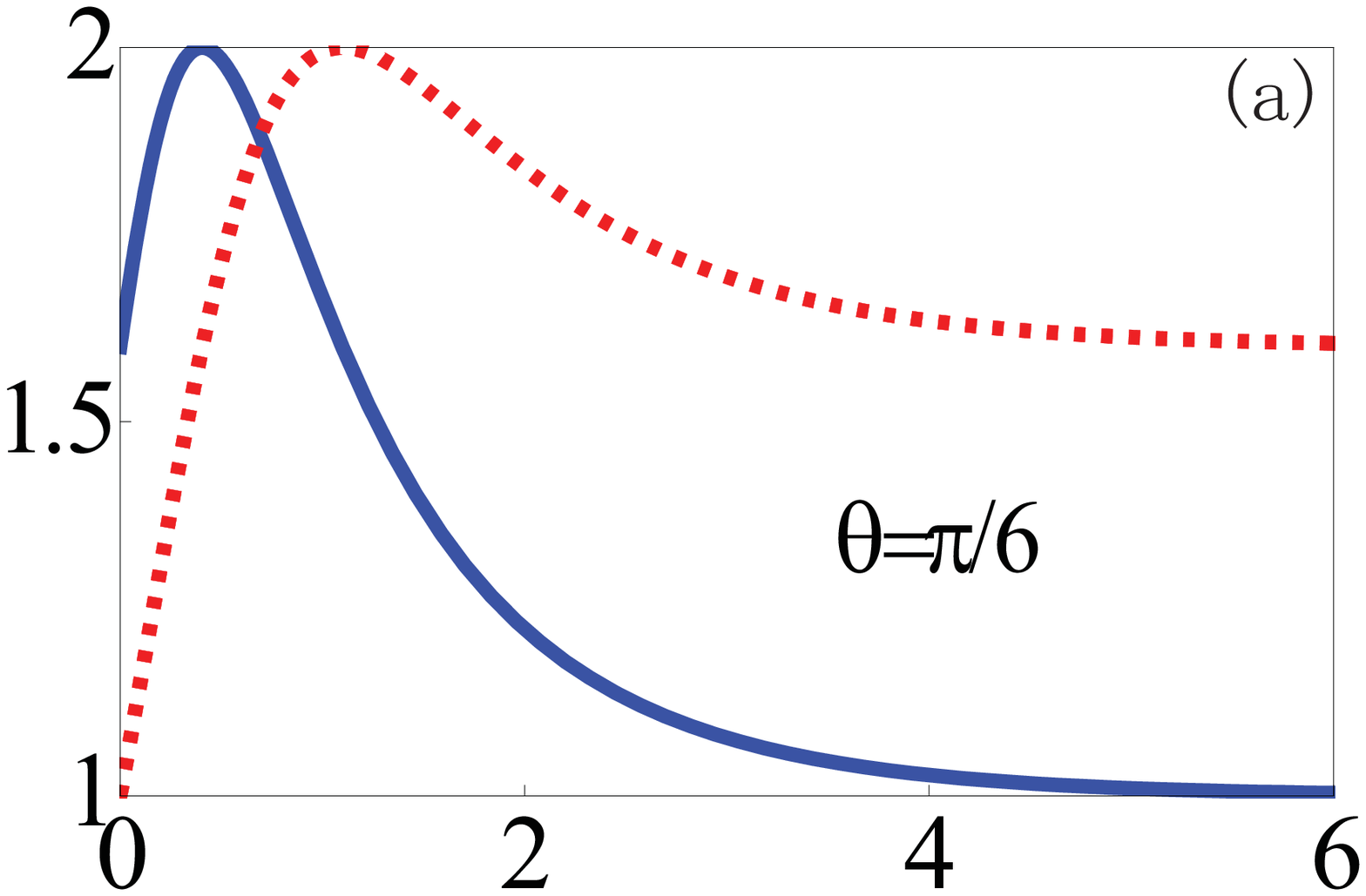}
\includegraphics[width= 4.2 cm]{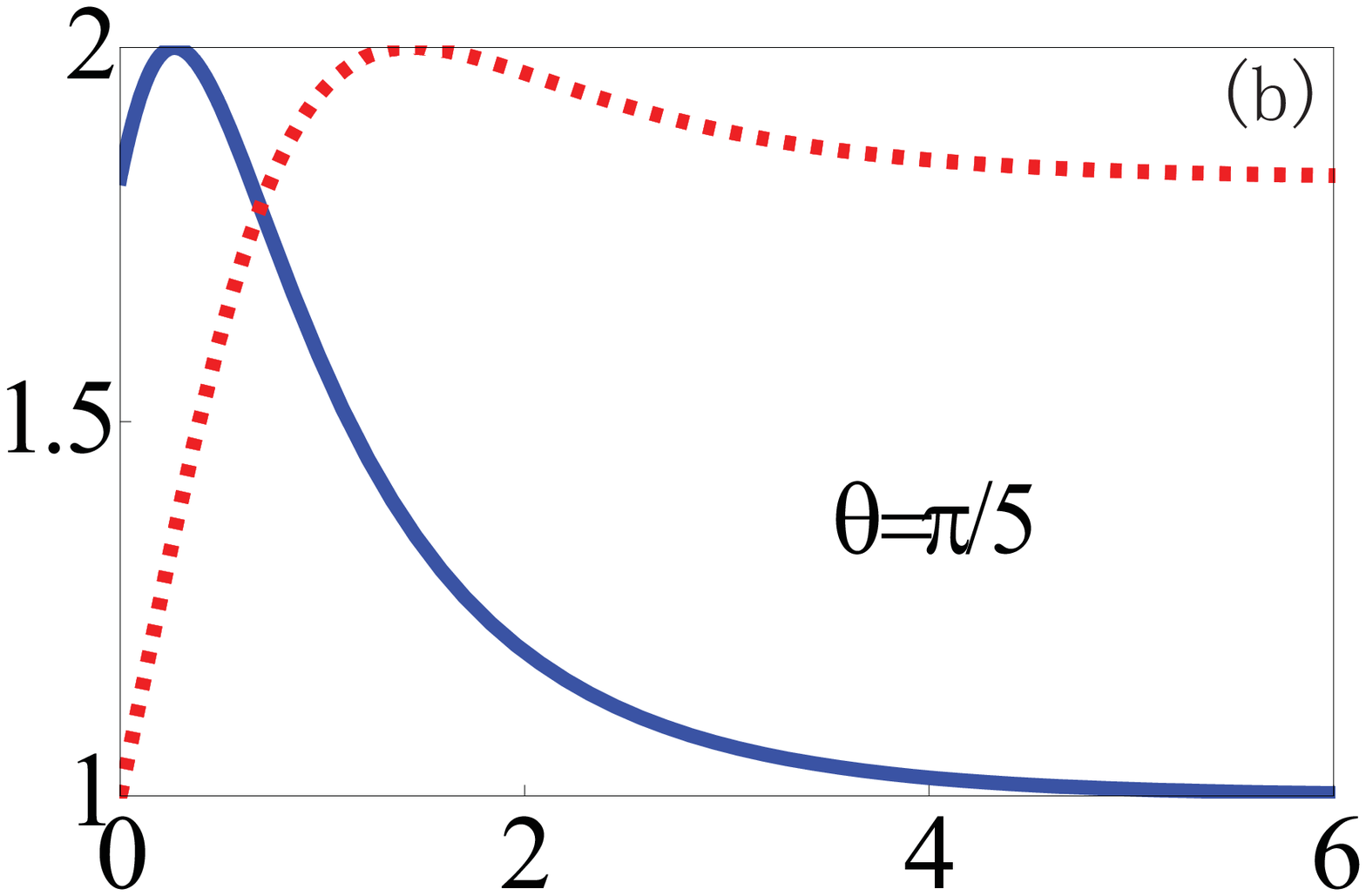}
\includegraphics[width= 4.2 cm]{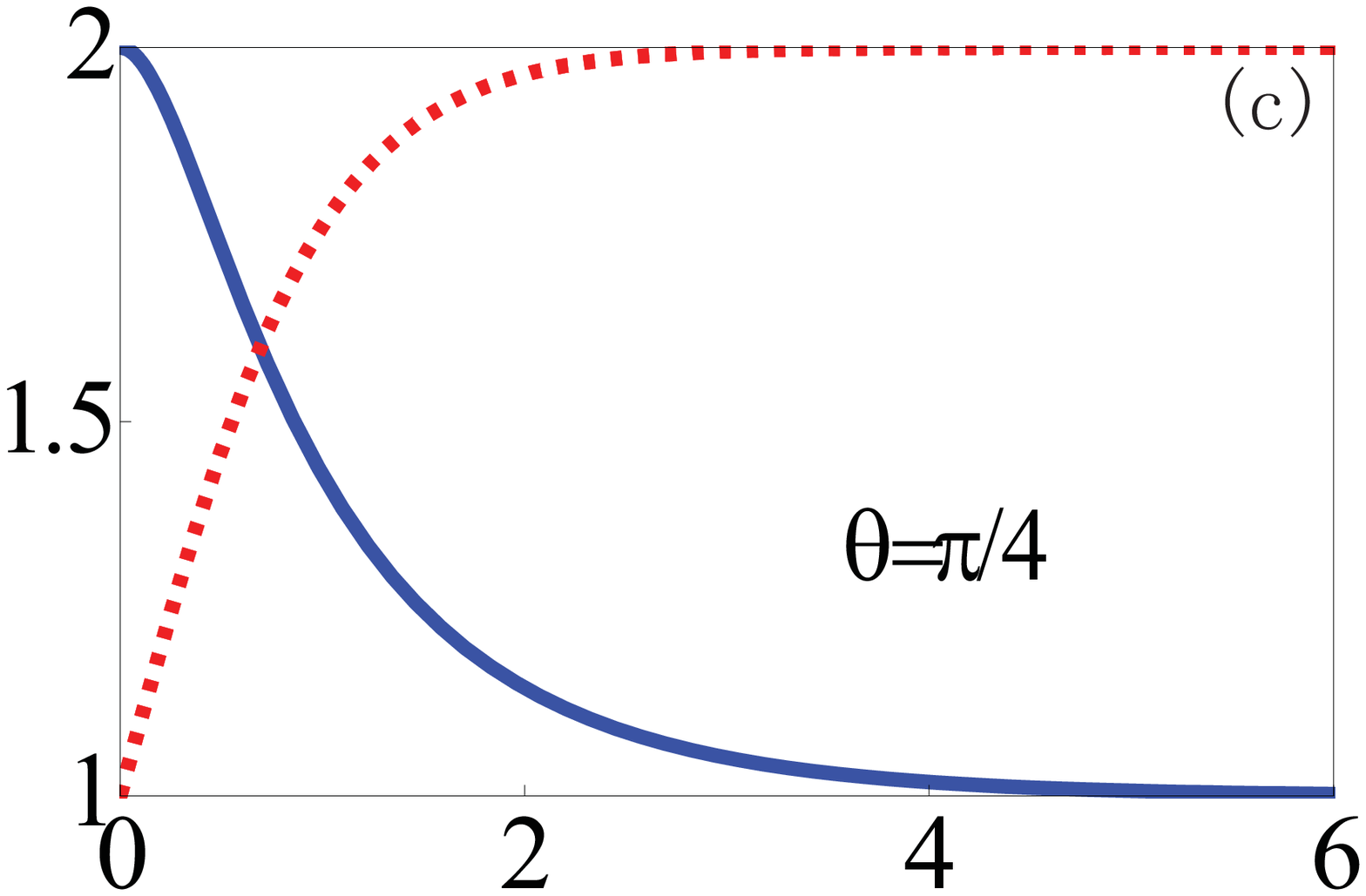}
\includegraphics[width= 4.2 cm]{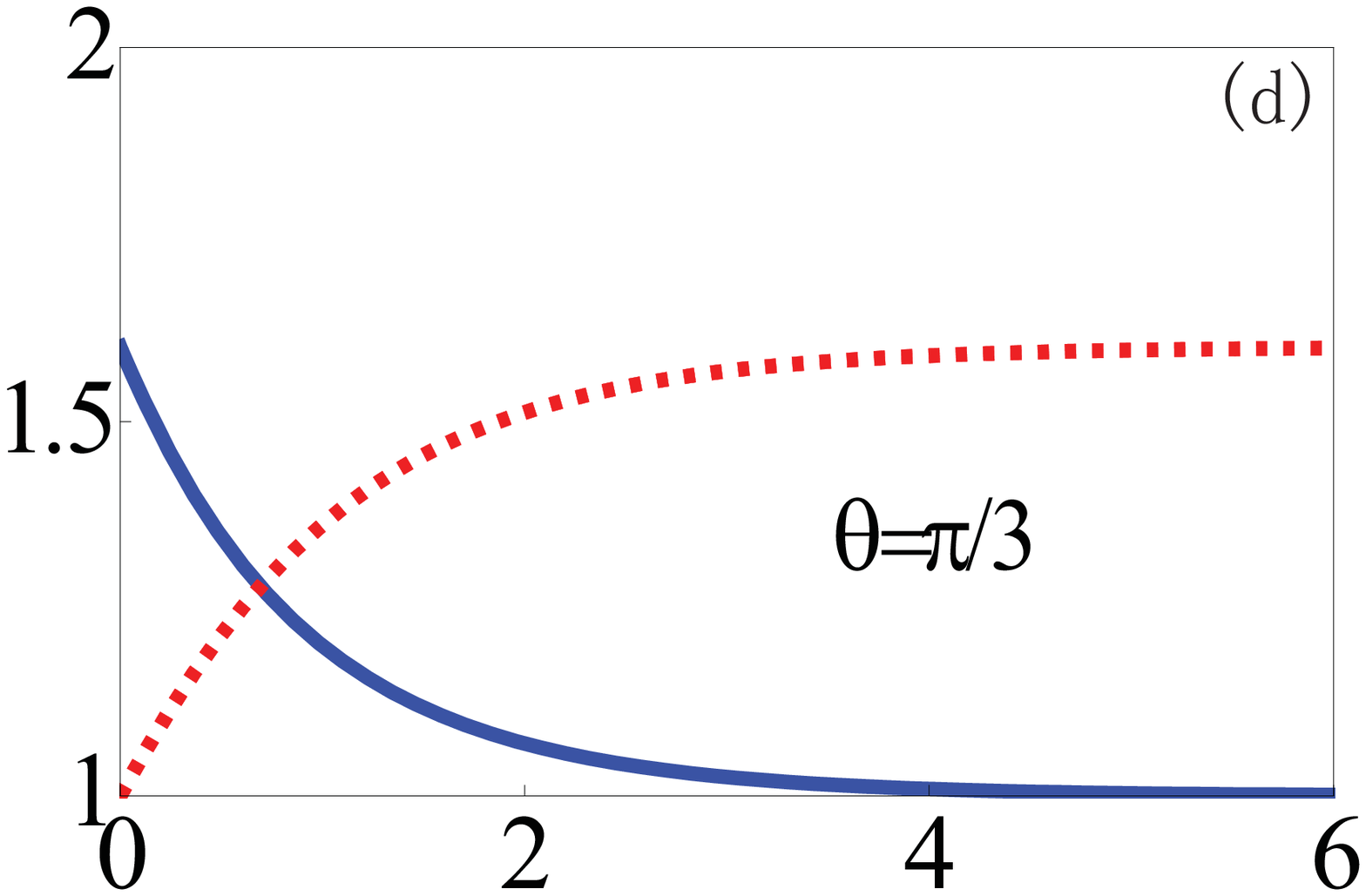}
\caption{Time dependence of Schmidt weights $K_{A}(t)$ and
$K_{a}(t)$ for the spontaneous emission dynamics at different values
of $\theta$. In each of the four plots, the blue solid and the red
dotted lines denote $K_{A}(t)$ and $K_{a}(t)$ respectively, and the
x axis represents time $t$ with unit $1/\Gamma _{A}$. Complementary
behavior of the two Schmidt weights is shown clearly in plots (c)
and (d) where $\sin^{2}\theta\ge \cos^{2}\theta$.} \label{figSE}
\end{figure}

We note that at $t=0$,
\begin{equation}
K_{A}(0) = \frac{1}{\cos^{4}\theta + \sin^{4}\theta},
\end{equation}
a finite number that is naturally the same as the constant Moon
entanglement (\ref{MoonEntanglement}). As time goes on, the
probability of the atom in the excited state $P_{e}(t)$ decays
gradually, and at $t=\infty$ the probability is completely
transferred to the ground state and leaves the atom in a product
state with its remainder system, which means eventually $A$ is
disentangled from the rest of the universe, and the Schmidt weight
$K_{A}(\infty )=1$. Fig. \ref{figSE} illustrates the behavior
$K_{A}(t)$ as a function of $t$ at four different $\theta$ values.
We note that in the region when $\sin^{2}\theta <\cos^{2}\theta$,
$K_{A}(t)$ starts from a finite value, evolves to a local maximum
and then decays irreversibly to $1$ as is shown in Fig. \ref{figSE}
(a) and (b). However when $\sin^{2}\theta\ge \cos^{2}\theta$ as is
shown in Fig. \ref{figSE} (c) and (d), $K_{A}(t)$ decays directly
and irreversibly to $1$.

Now let us focus on the reservoir entanglement. From the time
dependent state (\ref{Spon state}) we see that the coefficient
matrix of reservoir $a$ is given as
\begin{equation}
C_{a}=\left(
\begin{array}{cccc}
0 & \sin \theta  & c_{e}(t)\cos \theta  & ... \\
c_{1}(t)\cos \theta  & 0 & 0 & ... \\
.. & \vdots  & \vdots  & \vdots  \\
c_{k}(t)\cos \theta  & \vdots  & \vdots & \vdots  \\
.. & \vdots  & \vdots  & \vdots
\end{array}
\right).
\end{equation}
Then the reduced density matrix is given by
$\rho_{a}=C_{a}C_{a}^{\dag}$ with an infinite number of non-zero
eigenvalues
\begin{eqnarray}
{\lambda _{0}} &{=}&|c_{e}(t)|^{2}\cos ^{2}\theta +\sin ^{2}\theta, \\
{\lambda _{k}} &{=}&|c_{k}(t)|^{2}\cos ^{2}\theta,
\end{eqnarray}
for $k=1,2,3,...,\infty$. Now from the definition (\ref{Schmidt}) we
find the reservoir Schmidt weight:
\begin{equation}
K_{a}(t)=\frac{2}{[2\cos ^{2}\theta (1-e^{-\Gamma _{A}t})-1]^{2}+
1}. \label{SponK_a}
\end{equation}
Obviously the reservoir is initially not entangled. Then its
entanglement gradually increases, and at time $t=(\ln 2)/\Gamma
_{A}$ we find $K_{A}(t)=K_{a}(t)$. When time goes to infinity we
note
\begin{equation}
K_{a}(\infty ) = K_{A}(0) = \frac{1}{\cos^{4}\theta +
\sin^{4}\theta}.
\end{equation}
That is, the final reservoir entanglement $K_{a}(\infty )$ equals
the initial qubit entanglement $K_{A}(0)$. Fig. \ref{figSE} plots
the behavior of $K_{a}(t)$ as a function of $t$ for various $\theta
$ values. We note that in the region when $\sin^{2}\theta
<\cos^{2}\theta$ as is shown in Fig. \ref{figSE} (a) and (b),
$K_{a}(t)$ starts from zero entanglement, reaches a maximum and then
evolves to a finite value $K_{A}(0)$ in the end. In the opposite
region of $\theta$ as shown in Fig. \ref{figSE} (c) and (d), it
increases directly and irreversibly to the value $K_{A}(0)$.

To summarize, the qubit entanglement $K_{A}(t)$ starts at a finite
value and decays completely to zero entanglement, while $K_{a}(t)$
starts from no entanglement and eventually inherits the exact amount
of the qubit's initial entanglement $K_{A}(0)$. This is exactly
equal to the Moon entanglement $K_{M}$, so one can see that the
unknown Moon's entanglement (\ref{MoonEntanglement}) has jumped into
the picture. It is constant itself, but it acts as a kind of buffer
to restrict information flow to and from $A$. In another way of
speaking, we could say that there is only a certain amount of
``free" entanglement able to be exchanged, which is determined by
the Moon.

To take a further step and without loss of generality, we now assume
$\sin ^{2}\theta \geq \cos ^{2}\theta $ for convenience. Then we
find two equalities connecting each of $K_{A}(t)$ and $K_{a}(t)$ to
$K_{M}$:
\begin{equation}
\sqrt{\frac{2}{K_{A}(t)}-1}-\sqrt{\frac{2}{K_{M}}-1} =2(1-e^{-\Gamma
_{A}t})\cos ^{2}\theta,  \label{SponRestrict_A}
\end{equation}
\begin{equation}
\sqrt{\frac{2}{K_{a}(t)}-1}-\sqrt{\frac{2}{K_{M}}-1} =2e^{-\Gamma
_{A}t}\cos ^{2}\theta.  \label{SponRestrict_a}
\end{equation}
We note that both $K_{A}(t)$ and $K_{a}(t)$ are controlled by the
Moon in a non-linear way. This control leads to a novel conservation
relation between $A$ and $a$ in a pairwise fashion:
\begin{equation}
\sqrt{\frac{2}{K_{A}(t)}-1}+\sqrt{\frac{2}{K_{a}(t)}-1} =
1+\sqrt{\frac{2}{K_{M}}-1}. \label{SponInvariant}
\end{equation}
Although $K_{A}(t)$ and $K_{a}(t)$ are time dependent quantities
they combine in this way to a constant determined only by the Moon
entanglement $K_{M}$.

Because of the restriction of the Moon entanglement we note from the
conservation relation (\ref{SponInvariant}) that in this region
$\sin^{2}\theta \geq \cos^{2}\theta$ or $\theta \in [\pi/4,3\pi/4]$,
the decreasing of $K_{A}(t)$ is accompanied by the increasing of
$K_{a}(t)$ as is shown in Fig. \ref{figSE} (c) and (d). To see
quantitatively this complementary relation let us take
$\theta=\pi/4$ as an example. Then the time dependent qubit and
reservoir Schmidt weights simplify to
\begin{eqnarray}
K_{A}(t) &=&\frac{2}{(1-e^{-\Gamma_{A}t})^{2}+1},  \label{SponComplK_A} \\
K_{a}(t) &=&\frac{2}{e^{-2\Gamma_{A}t}+1}.  \label{SponComplK_a}
\end{eqnarray}
Obviously these two equations for $K_{A}(t)$ and $K_{a}(t)$ depend
on time in an opposite way. It is interesting to note that the
parameter $\Gamma_{A}$, which represents a collective coupling
between the atom and the reservoir modes, is also controlling the
two entanglements inversely. This is because of the Moon
entanglement $K_{M}$, which acts as a buffer to both entanglements
$K_{A}(t)$ and $K_{a}(t)$ but substantially in a opposite way
through $\Gamma _{A}$ (see Eqs. (\ref{SponRestrict_A}) and
(\ref{SponRestrict_a})).

We remark that for $\sin ^{2}\theta <\cos ^{2}\theta$, a similar
relation to Eq. (\ref{SponInvariant}) can be achieved with only a
modification of signs. In this case $K_{A}(t)$ and $K_{a}(t)$ are
not always complementary any more as Fig. \ref{figSE} (a) and (b)
show for wide regions of $t$. However $K_{A}(t) $ and $K_{a}(t)$
still stand in a time-invariant relation similar to
(\ref{SponInvariant}) and are connected only by $K_{M}$.

%+++++++++++++++++++++++++++++++++++++++++++++
\section{Jaynes-Cummings interaction}

Spontaneous emission is an example of an irreversible process. In
this section we will turn to a simple example when the $A$-$a$
interaction is reversible, following the JC model \cite{JC}. Thus
qubit $A$ is still a two level atom while unit $a$ is now simply a
single mode lossless cavity. Local entanglement dynamics between the
atom and the field in the JC model was first studied by Phoenix and
Knight \cite{Phoenix-Knight} by expressing the entangled atom-field
state in terms of the eigenstates and eigenvalues of both the field
and atomic operators, and revival physics \cite{Eberly-etal80,
Eberly-etal81, Yoo85} played a key role in the dynamics. Later, it
was shown by Son, et al., \cite{Son-etal02} that entanglement
between two non-interacting qubits can be generated through the
qubits' local interactions with their corresponding JC cavities that
are initially in an entangled two-mode squeezed state. Lee, et al.,
\cite{Kim-etal} showed that there is actually entanglement
reciprocation between the two qubits and their corresponding
continuous-variable systems such as JC cavities. Recently, the JC
model was revisited by Y\"{o}na\c{c}, et al., \cite{Yonac-etal06,
Yonac-etal07, Yonac-Eberly08}, and by Sainz and Bj\"{o}rk
\cite{Sainz-Bjork07}, to illustrate the entanglement sudden death
phenomenon \cite{Yu-Eberly04}, as well as to track the entanglement
flow, and conservation relations were found by both groups
\cite{Yonac-etal07, Sainz-Bjork07}.

Here we will continue to track the entanglement information in the
JC dynamics, but in addition will account quantitatively for the
role of the non-interacting unknown Moon. The JC Hamiltonian is
given as
\begin{equation}
H_{Aa}=\frac{1}{2}\hbar \omega _{A}\sigma _{A}^{z}+\hbar g(a^{\dag
}\sigma _{A}^{-}+a\sigma_{A}^{+})+\hbar \omega _{a}a^{\dag }a,
\end{equation}
where $\sigma _{A}^{z}$, $\sigma _{A}^{\pm }$ are the usual Pauli
matrices describing the two level atom $A$, while $a^{\dag }$ and
$a$ denote the standard creation and annihilation operators for the
single mode cavity. The atom and cavity frequencies are $\omega
_{A}$ and $\omega _{a}$, respectively, and $g$ is the coupling
constant between the atom and the cavity. For convenience we take
the resonant condition when $\omega _{A}=\omega _{a}$.

Now from the generic expression (\ref{initial}) the initial state
for the JC model can be written as
\begin{equation}
|\psi _{AaM}(0)\rangle =\Big(\cos \theta |e\rangle |m_{1}\rangle
+\sin \theta |g\rangle |m_{2}\rangle \Big)\otimes |0\rangle,
\end{equation}
where $|e\rangle $ and $|g\rangle $ are the excited and ground state
of the two level atom, and we have defined $|\phi _{1}\rangle
=|0\rangle $ as the zero photon state of the cavity. From the
Jaynes-Cummings treatment \cite{JC} we will have the time dependent
state as
\begin{eqnarray}
|\psi _{AaM}(t)\rangle &=&\cos\theta|m_{1}(t)\rangle
\Big(e^{i\omega_{A}t/2\hbar}\cos gt|e\rangle |0\rangle   \notag \\
&-& ie^{-i\omega _{A}t/2\hbar }\sin gt|g\rangle |1\rangle \Big)  \notag \\
&+& \sin \theta |m_{2}(t)\rangle |g\rangle |0\rangle,
\end{eqnarray}
where $|1\rangle $ means that there is one photon in the cavity. If
we follow the same Schmidt calculations as in the last section, we
will have entanglement $K_{A}(t)$ between atom $A$ and the rest of
the universe as
\begin{equation}
K_{A}(t)=\frac{2}{[2\cos ^{2}\theta \cos ^{2}gt-1]^{2}+1}.
\label{JCK_A}
\end{equation}
That is, we find expression (\ref{SponK_A}) again, except that
$e^{-\Gamma_{A}t}$ has been replaced by $\cos ^{2}gt$. This is just
the replacement of one formula for excited state probability by
another, as the nature of the amplitude decay channel requires.

\begin{figure}[t!]
\includegraphics[width= 4.2 cm]{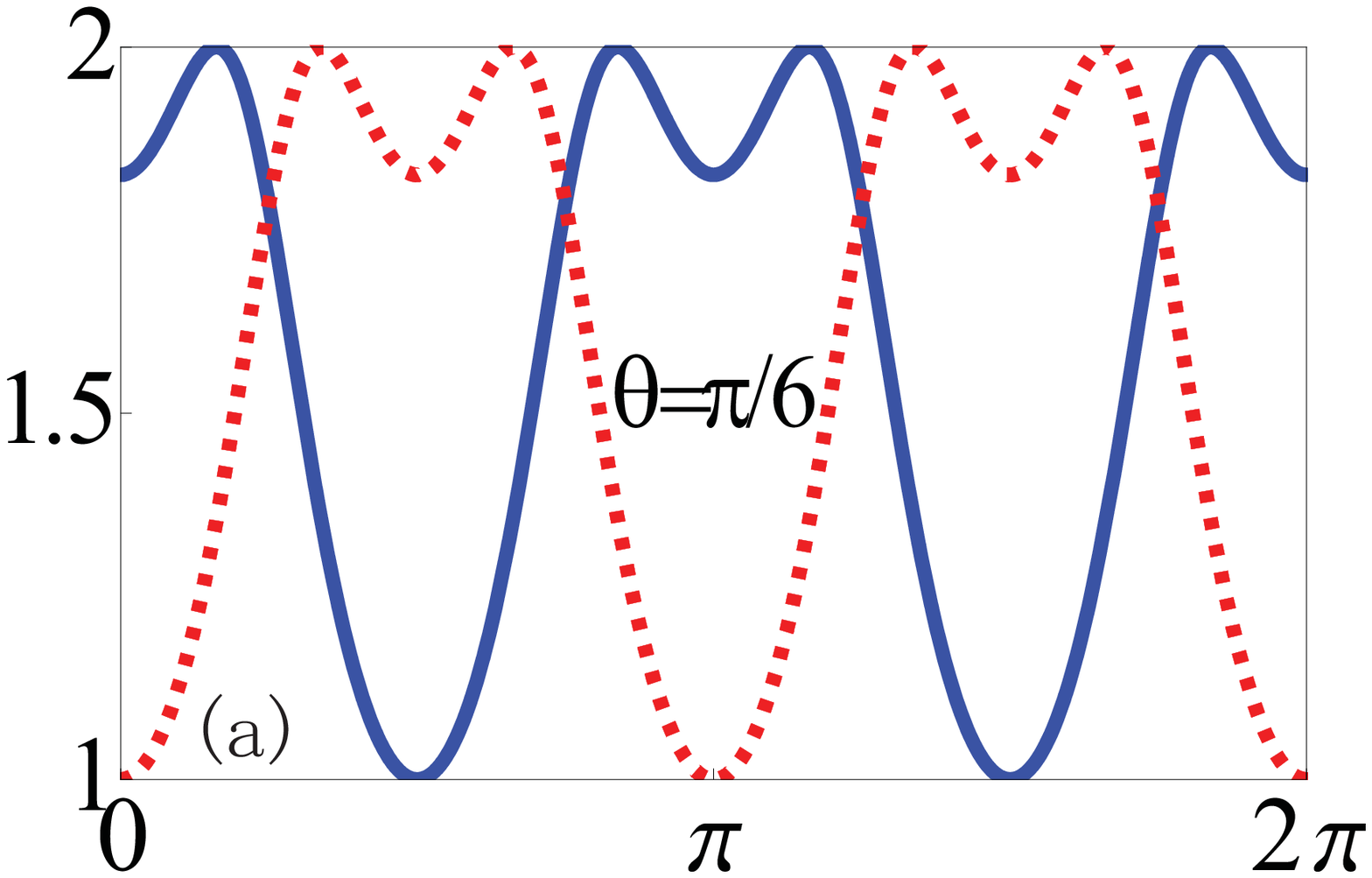}
\includegraphics[width= 4.2 cm]{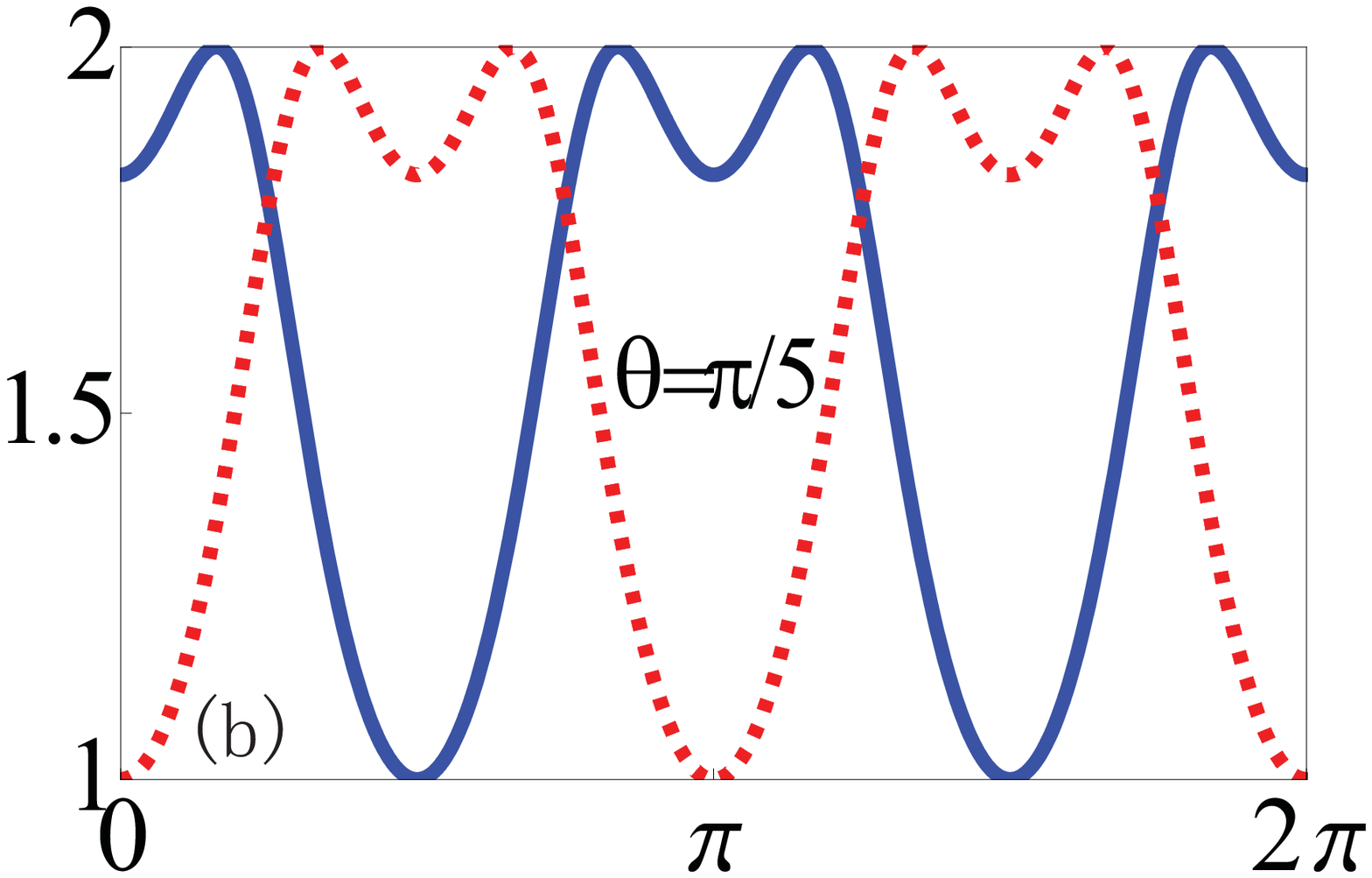}
\includegraphics[width= 4.2 cm]{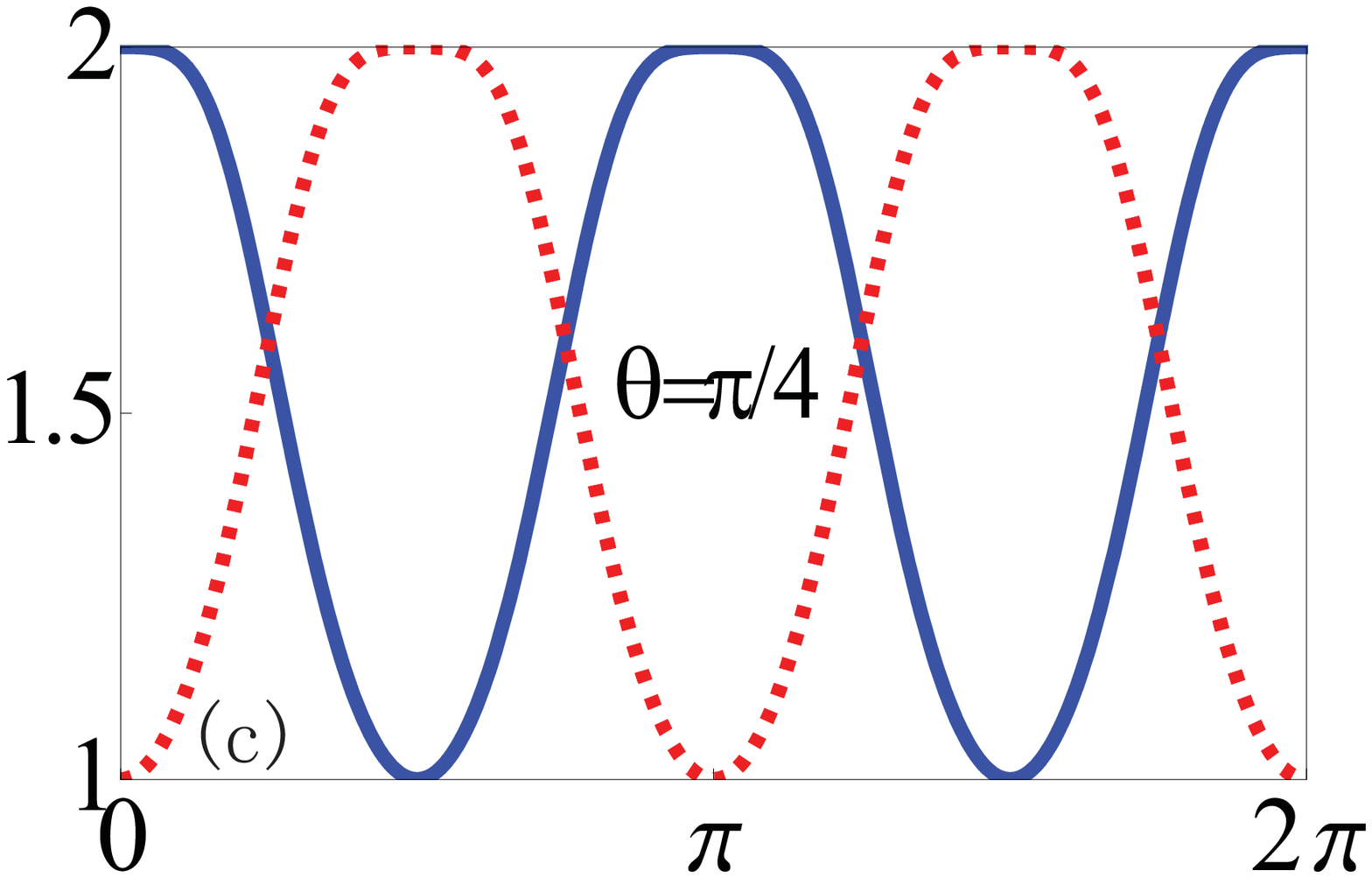}
\includegraphics[width= 4.2 cm]{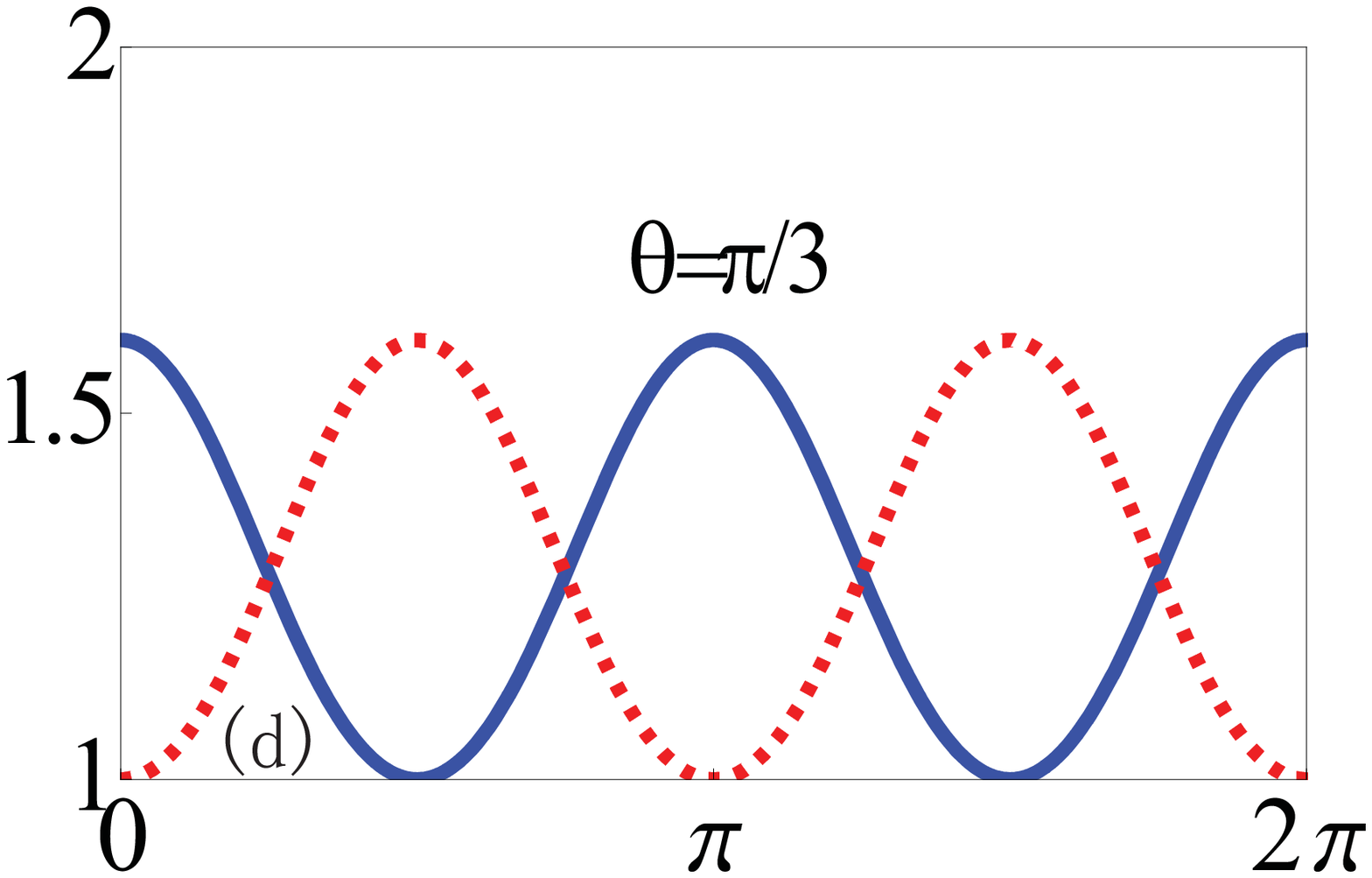}
\caption{Time dependence of Schmidt weights $K_{A}(t)$ and
$K_{a}(t)$ for the JC dynamics at different values of $\theta$. In
each of the four plots, the blue solid and the red dotted lines
denote $K_{A}(t)$ and $K_{a}(t)$ respectively, and the x axis
represents time $t$ with unit $1/g$. Complementary behavior of the
two Schmidt weights is shown clearly in plots (c) and (d) where
$\sin^{2}\theta\ge \cos^{2}\theta$.} \label{figJC}
\end{figure}

While the initial qubit entanglement $K_{A}(0)$ again has a value
equal to the Moon entanglement (\ref{MoonEntanglement}), in the JC
dynamics $K_{A}(t)$ has a period of $\tau =\pi /g$ instead of
decaying irreversibly as in the spontaneous emission case. We see at
the half period time the atom loses all of its entanglement:
$K_{A}(t=\tau /2)=1$. Then it evolves to the initial value
$K_{A}(0)$ at $t=\tau$. Fig. \ref{figJC} shows this periodic
behavior of $K_{A}(t)$ plotted as a function of $t$ at different
$\theta$ values. Recovery of atom-field and atom-atom entanglement
in the JC dynamics was already shown previously in Refs.
\cite{Phoenix-Knight, Bose-etal01, Yonac-etal06, Yonac-etal07,
Son-etal02, Sainz-Bjork07, Yonac-Eberly08}. However, here our result
shows a different type of entanglement recovery, because $K_{A}(t)$
denotes another type of entanglement, this time including the
unspecified non-interacting Moon as well as the cavity.

The cavity entanglement $K_{a}(t)$,
\begin{equation}
K_{a}(t)=\frac{2}{[2\cos ^{2}\theta \sin ^{2}gt-1]^{2}+1},
\label{JCK_a}
\end{equation}
is also predictable if we look to (\ref{SponK_a}) and see that
$1-e^{-\Gamma_{A}t}$ should be converted to $\sin^{2}gt$ because
both are expressions for the ground state probability. We note that
$K_{a}(t)$ is also periodic. It is initially not entangled with its
remainder ($K_{a}(0)=1$), and then increases with time. At $t=\pi
/4g$, we have $K_{a}(t)=K_{A}(t)$, and at the half period time we
see that
\begin{equation}
K_{a}(t=\tau /2)=K_{A}(0)=K_{M},
\end{equation}
exactly the same as $K_{A}(0)$ entanglement. Again Fig. \ref{figJC}
illustrates the periodic behavior of $K_{a}(t)$ as a function of $t$
at various $\theta $ values. When compared with the behavior of the
qubit entanglement we see that the amount of entanglement $K_{M}$
has been completely transferred from $K_{A}(t)$ to $K_{a}(t)$ at the
half period time $t=\tau /2$. After this, however, the entanglement
is repeatedly transferred back and forth between $K_{A}(t)$ and
$K_{a}(t)$. This is the major difference from the spontaneous
emission case where the reversible process is absent.

Again we work in the sector when $\sin ^{2}\theta \ge \cos
^{2}\theta$ for convenience and see that both of the entanglements
$K_{A}(t)$ and $K_{a}(t)$ are restricted by the constant Moon
entanglement $K_{M}$ in the following non-linear time-dependent way:
\begin{eqnarray}
\sqrt{\frac{2}{K_{A}(t)}-1}-\sqrt{\frac{2}{K_{M}}-1}
&=&2\sin ^{2}gt\cos ^{2}\theta,  \label{JCRestrict_A} \\
\sqrt{\frac{2}{K_{a}(t)}-1}-\sqrt{\frac{2}{K_{M}}-1} &=&2\cos
^{2}gt\cos ^{2}\theta.  \label{JCRestrict_a}
\end{eqnarray}
This periodic time dependent control of the two Schmidt weights by
the Moon entanglement is different from the spontaneous emission
case. However, the two equalities also lead to the same generic
entanglement conservation relation
\begin{equation}
\sqrt{\frac{2}{K_{A}(t)}-1}+\sqrt{\frac{2}{K_{a}(t)}-1}
=1+\sqrt{\frac{2}{K_{M}}-1}. \label{JCInvariant}
\end{equation}
Therefore in the JC model case the time dependent Schmidt weights
$K_{A}(t)$ and $K_{a}(t)$ are also restricted by the constant Moon
entanglement $K_{M}$. We see clearly here that the decrease of
$K_{A}(t)$ is accompanied by the increase of $K_{a}(t)$ and vice
versa as is shown in Fig. \ref{figJC} (c) and (d). To show
quantitatively we again take $\theta =\pi/4$ as in Fig. \ref{figJC}
(c) to follow this complementary relation of the two entanglements:
\begin{eqnarray}
K_{A}(t) &=&\frac{2}{\sin ^{4}gt+1},  \label{JCComplK_A} \\
K_{a}(t) &=&\frac{2}{\cos ^{4}gt+1},  \label{JCComplK_a}
\end{eqnarray}
which are the exact analogs of (\ref{SponComplK_A}) and
(\ref{SponComplK_a}).

%+++++++++++++++++++++++++++++++++++++++++++++
\section{XY Spin Interaction}

\begin{figure}[t!]
\includegraphics[width= 6 cm]{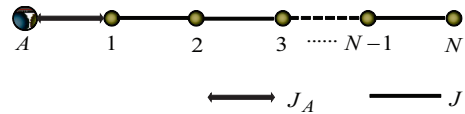}
\caption{Scheme of qubit $A$ interacting with an XY chain. The big
blue dot represents the qubit $A$ while the $N$ small dots represent
the spins in the XY chain. The arrow indicates interaction between
$A$ and the first spin in the chain with coupling constant $J_A$
while the solid segments connect the nearest neighbor sites in the
chain with coupling constant $J$.} \label{figXY0}
\end{figure}

We now move to a condensed matter context and take a final example
when the $A$-$a$ connection is a Heisenberg exchange interaction or
spin-spin interaction. Here qubit system $A$ is a spin one-half
particle while unit $a$ is now an $N$-spin XY chain \cite{Lieb61}
(see Fig. \ref{figXY0}), a simplified model for strongly correlated
materials such as ferromagnets, antiferromagnets, etc.

The first studies of entanglement flow in spin chains focused on
few-qubit chains ($N \le 6$) and the $W$ state \cite{Wang01}, and
also entanglement dispersion in long chains ($N \gg 1$)
\cite{Pratt-Eberly01}. Amico, et al. \cite{Amico-etal04} studied the
propagation of a pairwise entangled state through an XY spin chain,
and found that singlet-like states are transmitted with higher
fidelity than other maximally entangled states. Here we also focus
on the entanglement dynamics, not to transport the entanglement, but
to track the information flow by taking into account the role of the
entangled Moon. The interaction Hamiltonian of our scheme is given
as
\begin{eqnarray}
H_{Aa} &=&J_{A}(\sigma _{A}^{+}\sigma_{1}^{-}+\sigma_{A}^{-}\sigma_{1}^{+}) \notag \\
&+&\sum_{n=1}^{N-1}J(\sigma_{n}^{+}\sigma_{n+1}^{-}+\sigma_{n}^{-}\sigma_{n+1}^{+}),
\end{eqnarray}
where $\sigma^{\pm}$ are the usual Pauli matrices describing the
spins, $J_{A}$ is the coupling constant between spin $A$ and the
first spin $\sigma_{1}$\ of the XY chain, and $J$ is the coupling
constant between the nearest neighbor sites inside the XY chain. Now
we take $J_{A}=J$ for convenience. This model can be transformed
through a Jordan-Wigner transformation \cite{Jordan28} into a set of
free fermions (see for example Ref. \cite{Qian-Song06}) and thus can
be solved exactly \cite{Lieb61}. From the perspective of the
Jordan-Wigner transformation, the XY model is equivalent here to a
free fermion hopping model or Tight-Binding model describing phonon
systems.

For the XY Hamiltonian $H_{Aa}$ the exact $N+1$ eigenstates are
given as
\begin{eqnarray}
|k\rangle  &=&\sqrt{\frac{2}{N+2}}\sin \left(\frac{k\pi}{N+2}\right)
\left\vert \uparrow \right\rangle |0\rangle   \notag \\
&+&\sqrt{\frac{2}{N+2}}\sum_{n=1}^{N}\sin \left[\frac{\left(
n+1\right)k\pi}{N+2}\right] \left\vert\downarrow\right\rangle
|1_{n}\rangle,
\end{eqnarray}
where $\left\vert\uparrow\right\rangle$ and $\left\vert\downarrow
\right\rangle$ are the spin up and down states for our qubit $A$,
and $|1_{n}\rangle $, $|0\rangle $ are the states of the XY spin
chain with $|1_{n}\rangle $ indicating there is a spin up at the
$n$th site while all the rest are in the spin down state and
$|0\rangle $ meaning all the sites from site $1$ to site $N$ are in
the down state. Here $k=1,2,3,...,N+1$ represent the $N+1$
eigenstates. The corresponding eigenvalues are
\begin{equation}
E_{k}=2J\cos \left( \frac{k\pi }{N+2}\right).
\end{equation}
Then the evolution operator can be written as
\begin{equation}
U_{Aa}(t)=\sum_{k=1}^{N+1}e^{-iE_{k}t}|k\rangle \langle k|.
\label{XYUoperator}
\end{equation}

Again from the generic initial state (\ref{initial}) we have here
for the XY model
\begin{equation}
|\psi _{AaM}(0)\rangle =\Big(\cos \theta \left\vert\uparrow
\right\rangle |m_{1}\rangle +\sin \theta \left\vert\downarrow
\right\rangle |m_{2}\rangle \Big)\otimes |0\rangle ,
\end{equation}
where we have defined $|\phi _{1}\rangle =|0\rangle$ to represent
all the $N$ spins in the XY chain that are in the down state. Then
the time dependent state can be achieved as
\begin{eqnarray}
|\psi _{AaM}(t)\rangle  &=&\cos \theta
|m_{1}(t)\rangle\Big(c_{e}(t)|\uparrow \rangle |0\rangle
+\sum_{n=1}^{N}c_{n}(t)\left\vert \downarrow \right\rangle |1_{n}\rangle\Big)  \notag \\
&+&\sin \theta |m_{2}(t)\rangle |\downarrow \rangle |0\rangle.
\end{eqnarray}
where we have defined
\begin{equation}
c_{e}(t)=\sum_{k=1}^{N+1}\frac{2e^{-iE_{k}t}}{N+2}\sin \left(
\frac{k\pi}{N+2}\right) \sin \left( \frac{k\pi }{N+2}\right),
\end{equation}
\begin{equation}
c_{n}(t)=\sum_{k=1}^{N+1}\frac{2e^{-iE_{k}t}}{N+2}\sin \left(
\frac{k\pi}{N+2}\right) \sin \left( \frac{n+1}{N+2}k\pi \right).
\end{equation}

Since $c_{e}(t)$ and $c_{n}(t)$ are complicated expressions for
arbitrary number $N$, here we take $N=10$ as an example to
illustrate their properties. Then we have
\begin{eqnarray}
c_{e}(t)&=&\frac{1}{12}\Big[2+3\cos(Jt)+2\cos(\sqrt{2}Jt)+\cos(\sqrt{3}Jt) \notag \\
&+& (2+\sqrt{3})\cos\frac{(\sqrt{3}-1)Jt}{\sqrt{2}} \notag \\
&+& (2-\sqrt{3})\cos\frac{(\sqrt{3}+1)Jt}{\sqrt{2}} \Big].
\end{eqnarray}
We note that the five cosine functions have five different periods
and the ratio of any two periods is irrational. Therefore the five
quantities will not have a common period, which means that
$c_{e}(t)$ will oscillate all the time but without a fixed period.
Now we define $f(J,t)=|c_{e}(t)|^{2}$ and note that it can vary from
$0$ to $1$. There are infinitely many solutions for $f(J,t)=0$ as a
function of time $t$, say $t=\tau _{i}$, with $i=1,2,3,...,\infty $.

\begin{figure}[t!]
\includegraphics[width= 4.2 cm]{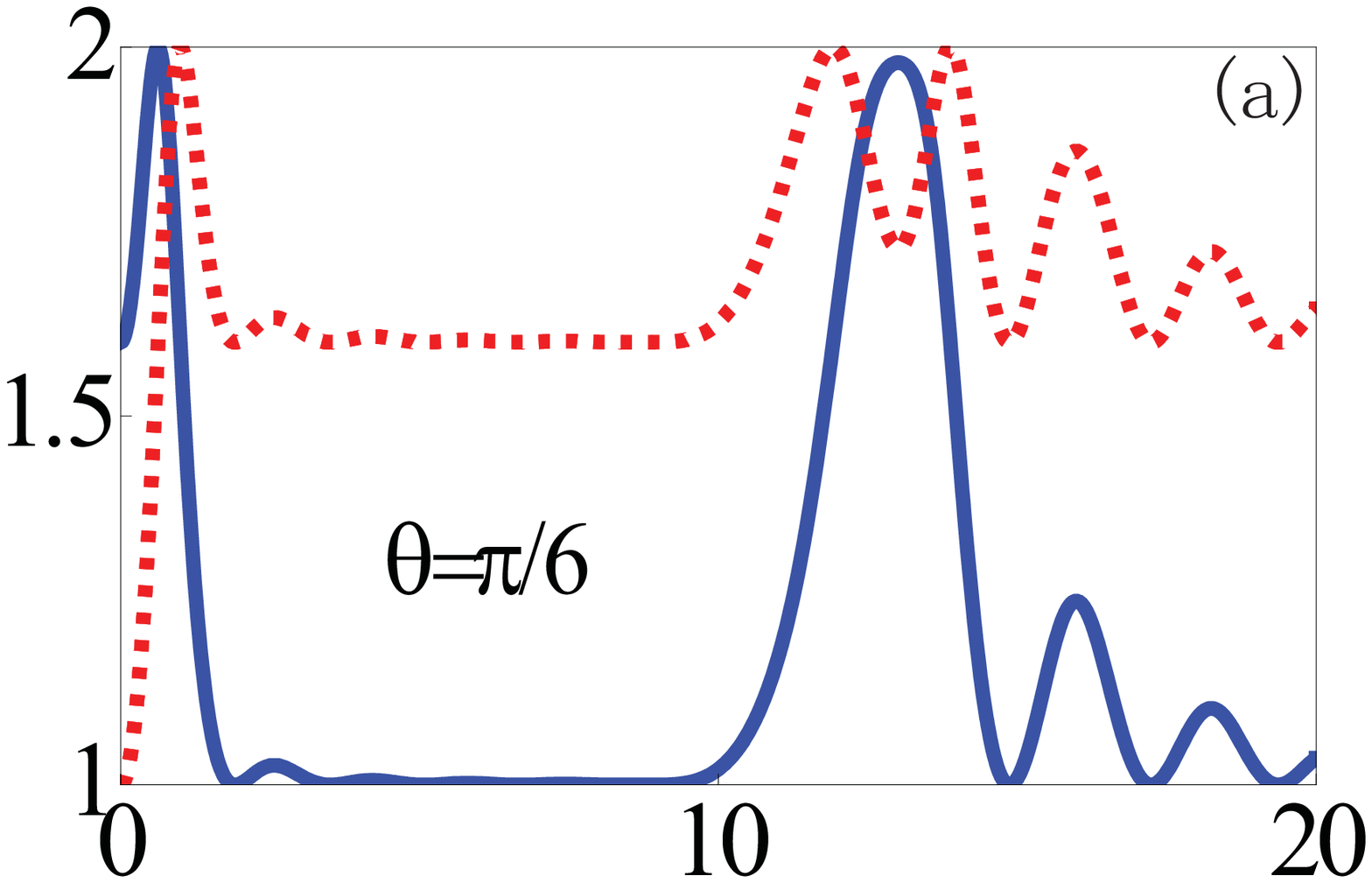}
\includegraphics[width= 4.2 cm]{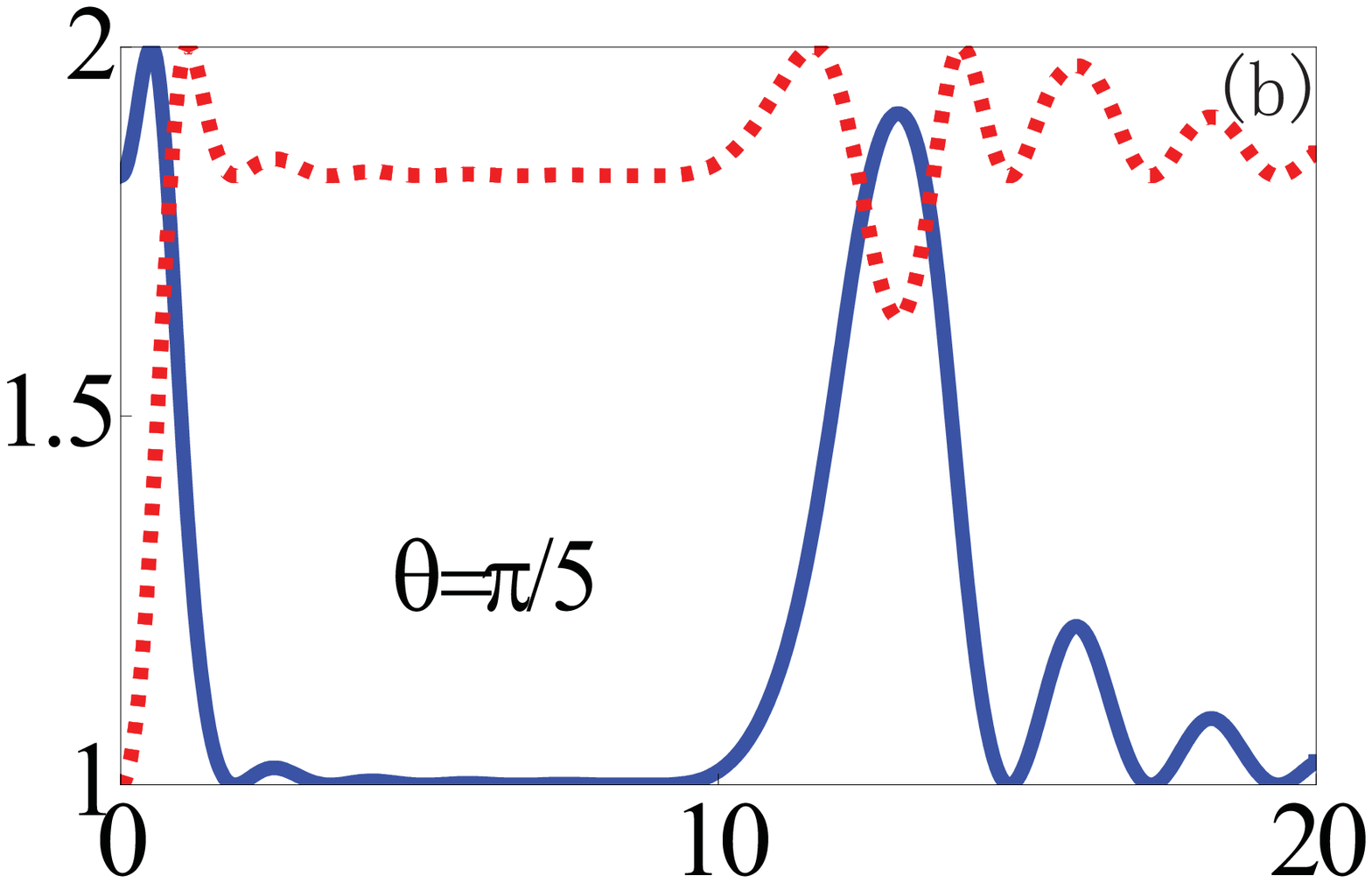}
\includegraphics[width= 4.2 cm]{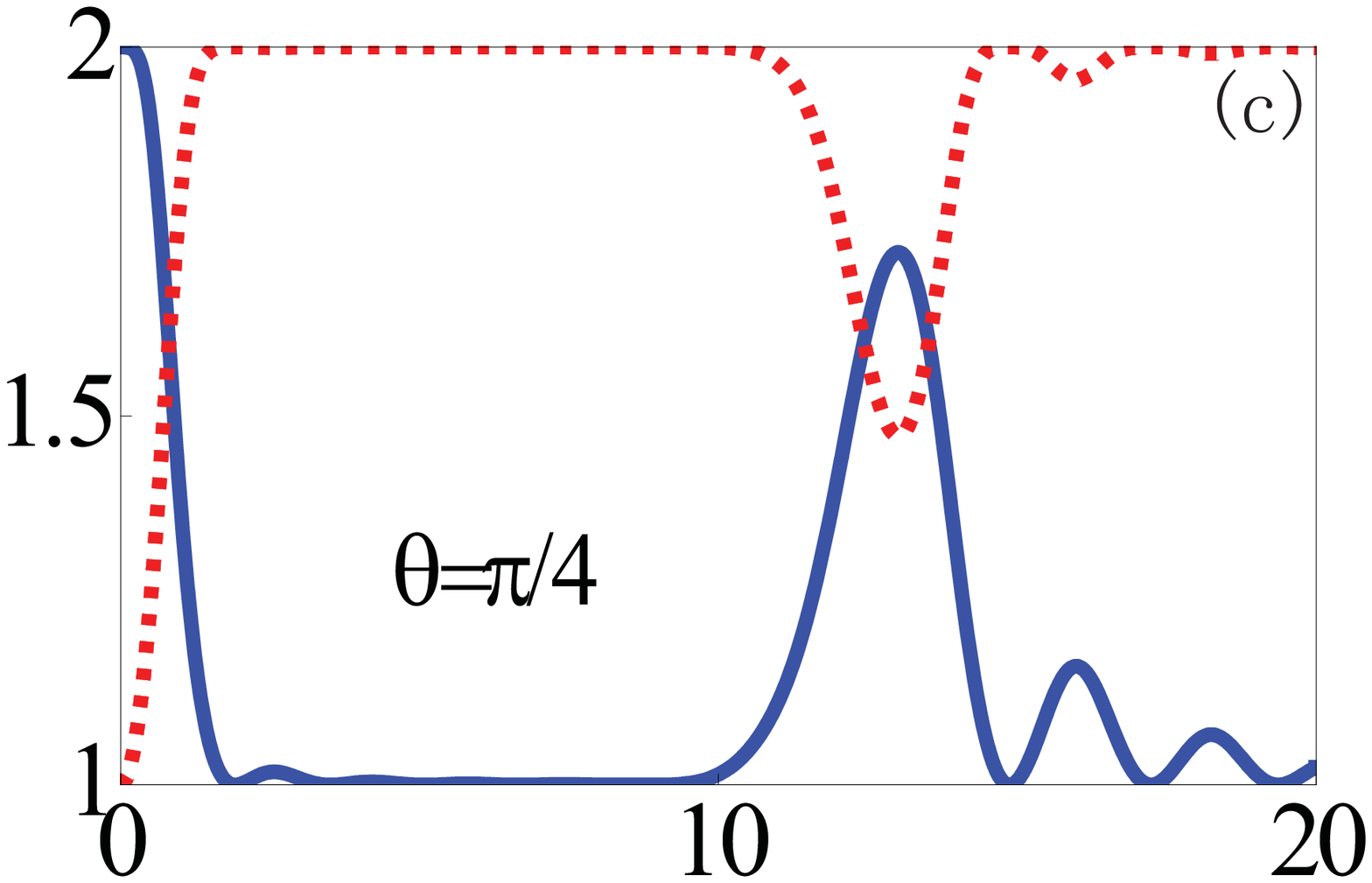}
\includegraphics[width= 4.2 cm]{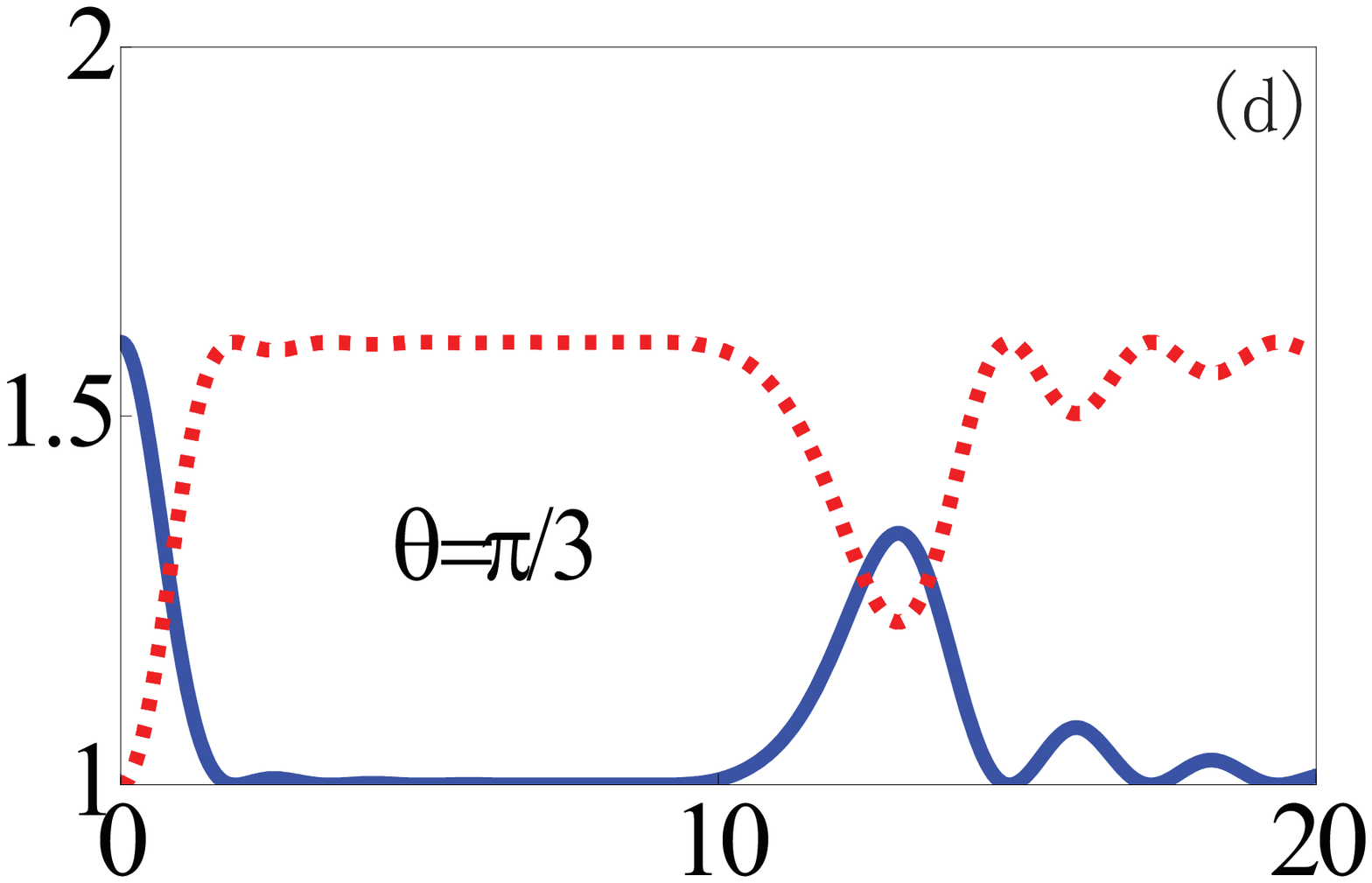}
\caption{Time dependence of Schmidt weights $K_{A}(t)$ and
$K_{a}(t)$ for the XY spin chain dynamics at different values of
$\theta$. In each of the four plots, the solid and the dotted lines
denote $K_{A}(t)$ and $K_{a}(t)$ respectively, and the x axis
represents time $t$ with unit $1/J$. Complementary behavior of the
two Schmidt weights is shown clearly in plots (c) and (d) where
$\sin^{2}\theta\ge \cos^{2}\theta$.} \label{figXY}
\end{figure}

If we follow the same Schmidt calculations as in the last two
sections we will find the Schmidt weight $K_{A}(t)$ between the
qubit spin $A$ and the remainder as
\begin{equation}
K_{A}(t)=\frac{2}{[2f(J,t)\cos^{2}\theta-1]^{2}+1}. \label{XYK_A}
\end{equation}
We note that the qubit entanglement $K_{A}(t)$ is also oscillating
as determined by $f(J,t)$. As the amplitude channel requires, it
starts at the familiar same value $K_{A}(0)=K_M$, and in this
example evolves to zero entanglement at the time points $\tau _{i}$.
After each of these zeros, $K_{A}(t)$ will increase to a local
maximum point and then decay to $1$ again at the next time point
$\tau_{i+1}$. Fig. \ref{figXY} illustrates this particular behavior
of $K_{A}(t)$ as a function of $t$ at different values of $\theta$.
Such aperiodic behavior is intermediate to the previous two examples
showing irreversible decay and periodic oscillation, and is expected
on the basis of the irrationally related spin-chain
eigenfrequencies.

Now we come to the XY chain entanglement $K_{a}(t)$ representing the
entanglement between the chain and its remainder, i.e., the end spin
$A$ and the Moon $M$. It is related to $K_A$ in the usual way. We
just replace $f(J,t)$ by $1-f(J,t)$ and obtain:
\begin{equation}
K_{a}(t)=\frac{2}{[2\cos ^{2}\theta(1 -f(J,t)) -1]^{2}+ 1}.
\label{XYK_a}
\end{equation}
So the chain entanglement also oscillates with $f(J,t)$. In general,
the entanglement will be transferred back and forth between
$K_{A}(t)$ and $K_{a}(t)$ with $K_{M}$ the upper limit of
entanglement that can be transferred just as the previous two cases.

As expected, restrictions on entanglement flow follow the previous
examples. In the regime $\sin ^{2}\theta \geq \cos ^{2}\theta$ we
can simply repeat relations (\ref{JCRestrict_A}) and
(\ref{JCRestrict_a}) by replacing $\cos^2gt$ with $f(J,t)$:
\begin{equation}
\sqrt{\frac{2}{K_{A}(t)}-1}-\sqrt{\frac{2}{K_{M}}-1}=
2\left[1-f(J,t)\right] \cos ^{2}\theta,  \label{XYRestric_A}
\end{equation}
\begin{equation}
\sqrt{\frac{2}{K_{a}(t)}-1}-\sqrt{\frac{2}{K_{M}}-1} =2f(J,t)\cos
^{2}\theta,  \label{XYRestric_a}.
\end{equation}
Naturally, the same non-linear conservation relation
(\ref{JCInvariant}) is recovered, and again the $A$ and $a$
entanglements behave complementarily, this time as a function of
$Jt$ as shown in Fig. \ref{figXY} (c) and (d).

%+++++++++++++++++++++++++++++++++++++++++++++
\section{Summary}

In summary we have studied entanglement information flow from the
perspective of a dynamical qubit $A$ in an initially mixed state, a
state that was generated by an entanglement associated with a prior
process, which we can loosely assign to an experimental preparation
stage.   Using Schmidt-decomposition rather than master-equation
analysis, we derived conservation statements for the separate
degrees of quantum entanglement of the qubit and of its interacting
reservoir, and showed their relation to the entanglement of the
unspecified background party we called the Moon, which was initially
entangled but at $t = 0$ ceased to interact with either the qubit
$A$ or its environment $a$.

The new forms of entanglement conservation relations are nonlinear
connections between quantum memories, dependent on the restrictions
implied by amplitude flow channel dynamics. One can say that the
channel's enforcement of excitation number conservation in the
qubit-reservoir interaction is the root cause of the entanglement
and its flow. This is closely analogous to the continuous
entanglement between transverse momenta in spontaneous parametric
down conversion, which arises from the enforcement of simultaneous
momentum and energy conservation on the two-photon amplitude in the
creation of the signal and idler photons.

Although unspecified, and ignored in previous open system analyses,
the Moon can be assigned responsibility for the initial impurity of
the qubit state. The three-part total universe ($A$ + $a$ + $M$) was
bi-partitioned three ways in order to evaluate the respective
Schmidt weights, as indicators of entanglements in three specific
interaction models (spontaneous emission, JC interaction, and XY
spin chain). These were analyzed to illustrate the flow of quantum
information in different contexts, including both discrete and
continuous versions of the reservoir system labelled $a$. Although
the influences on individual entanglements differ in various ways,
the amplitude flow common to them produces entanglement conservation
relations in the same form. One can say that the non-specified Moon
retains a kind of influence on the system of interest whether we are
``looking" (through interaction) at it or not. The qubit can feel,
through the entanglement conservation relation but not through
interaction, that the Moon is there.

There can be interesting consequences when the Moon also has a
significant dynamical evolution, although still not interacting with
$A$, because its entanglement with $A$ can then be assigned to part
rather than all of it. This discussion will be undertaken elsewhere
\cite{Qian-Eberly10}. Finally we would like to comment on the
inverse dependence of $K_{A}(t)$ and $K_{a}(t)$ on the interaction
parameters as discussed at the end of our three examples. It will be
particularly interesting if, for some systems, the interaction
constant can be adjustable (e.g., the coupling constant of a
spin-spin interaction). Especially in the thermodynamic limit
interesting phenomena such as quantum phase transitions
\cite{Sachdev, Osterloh-etal02} may arise from changes of the
interaction parameter. The behavior of the entanglements in the
vicinity of the critical point will be extremely interesting (see
for example \cite{Qian-etal05} and references therein).

%+++++++++++++++++++++++++++++++++++++++++++++
We acknowledge helpful conversations with Profs. L. Davidovich and
Ting Yu, and partial financial support from the following: DARPA
HR0011-09-1-0008, ARO W911NF-09-1-0385, NSF PHY-0855701.

%\newpage

\end{document}